\input{aipcheck}

\documentclass[
    ,final            % use final for the camera ready runs
  ]
  {aipproc}

\layoutstyle{8x11double}

\usepackage{graphicx}
\usepackage{epsfig}
\usepackage{latexsym}
\usepackage{amsmath}
\usepackage{amsfonts}
\usepackage{amsxtra}
\usepackage{axodraw}

\def\krto{ {\,\,\lower .8ex\hbox {$\longrightarrow \atop k \rightarrow 0$}\,\,}}

\def\bea{\begin{eqnarray} }
\def\beq{\begin{eqnarray} }

\def\eea{\end{eqnarray}}
\def\eeq{\end{eqnarray}}
\def\eq#1{eq.~(\ref{#1})}

\begin{document}

\title{The scaling infrared DSE solution as a critical end-point for the family of decoupling ones}

\classification{\texttt{12.35.Aw, 12.38.Lg, 12.38.Gc}}

\author{J. Rodr\'{\i}guez-Quintero}{
  address={Dpto. F\'isica Aplicada, Fac. Ciencias Experimentales; 
Universidad de Huelva, 21071 Huelva; Spain}
}

%\author{<author2>}{
%  address={<common address for author2 and author3>}
%}

%\author{<author3>}{
%  address={<common address for author2 and author3>}
%  ,altaddress={<author1 address>} % additional visiting address
%}

\begin{abstract}

Both regular (the zero-momentum 
ghost dressing function not diverging), also named decoupling, and 
critical (diverging), also named scaling, Yang-Mills propagators solutions 
can be obtained by analyzing the low-momentum behaviour of the ghost propagator Dyson-Schwinger 
equation (DSE) in Landau gauge. The asymptotic expression obtained for the 
regular or decoupling ghost dressing function up to the order ${\cal O}(q^2)$ fits pretty well 
the low-momentum ghost propagator obtained through the numerical integration of 
the coupled gluon and ghost DSE in the PT-BFM scheme. Furthermore, when the size of the 
coupling renormalized at some scale approaches some critical value, the PT-BFM results
seems to tend to the the scaling solution as a limiting case. This critical value of the coupling 
is compared with the lattice estimate for the Yang-Mills QCD coupling and the latter is shown 
to lie much above the former.

\end{abstract}

\maketitle

%%%%%%%%%%%%%%%%%%%%%%%%%%%%%%%%%%%%%%%%%%%%
%% MAINMATTER
%%%%%%%%%%%%%%%%%%%%%%%%%%%%%%%%%%%%%%%%%%%%

\section{Introduction}

The low-momentum behaviour of the Yang-Mills propagators derived either from 
the tower of Dyson-Schwinger equations (DSE) or from Lattice simulations in 
Landau gauge has been a very interesting and hot topic for the 
last few years. It seems by now well established that, if we assume 
in the vanishing momentum limit a ghost dressing function behaving as 
$F(q^2) \sim (q^2)^{\alpha_F}$ and a gluon propagator as 
$\Delta(q^2) \sim (q^2)^{\alpha_G-1}$ (or, by following a notation commonly used,
a gluon dressing function as $G(q^2)= q^2 \Delta(q^2) \sim (q^2)^{\alpha_G}$), 
two classes of solutions may emerge (see, for instance, the discussion 
of refs.~\cite{Boucaud:2008ji,Boucaud:2008ky}) from the DSE:
(i) those, dubbed {\it ``decoupling''}, where $\alpha_F=0$ and the suppression of 
the ghost contribution to the gluon propagator DSE results in a massive gluon 
propagator (see \cite{Aguilar:2006gr,Aguilar:2008xm} and references therein); 
and (ii) those, dubbed {\it ``scaling''}, where $\alpha_F \neq 0$ 
and the low-momentum behaviour of both gluon and ghost propagators 
are related by the coupled system of DSE through the condition $2 \alpha_F+\alpha_G = 0$ 
implying that $F^2(q^2)G(q^2)$ goes to a non-vanishing constant when $q^2 \to 0$ 
(see \cite{Alkofer:2000wg,Fischer:2008uz} and references therein). 

Lattice QCD results appear to support only the massive gluon ($\alpha_G=1$) or scaling solutions  
(see~\cite{Cucchieri:2007md} %,Bogolubsky:2007ud,IlgenGrib,Boucaud:2005ce,Oliveira:2010xc,Bornyakov:2009ug} 
and references therein), and also pinching technique results (see, for instance, 
\cite{Cornwall,Binosi:2002ft} and references therein), 
refined Gribov-Zwanziger~\footnote{In addition, K-I. Kondo triggered very recently an interesting 
discussion about the Gribov horizon condition and its implications on the Landau-gauge 
Yang-Mills infrared solutions~\cite{Kondo:2009ug,Dudal:2009xh,Aguilar:2009pp,Boucaud:2009sd}.} 
formalism (see~\cite{Dudal:2007cw}) or other approaches like the infrared mapping of $\lambda \phi^4$ and Yang-Mills 
theories in ref.~\cite{Frasca:2007uz} or the massive extension of the Fadeev-Popov 
action in ref.~\cite{Tissier:2010ts} appear to point to.

In the present contribution, we will briefly review the work of ref.~\cite{RodriguezQuintero:2010xx}, 
which extended the previous studies of refs.~\cite{Boucaud:2008ji,Boucaud:2008ky,Boucaud:2010gr}, 
by the analysis of the results obtained by solving the coupled system of Landau gauge ghost and gluon propagators DSE 
within the framework of the pinching technique in the background field method~\cite{Binosi:2002ft} (PT-BFM)

\section{The two kinds of solutions of the ghost propagator Dyson-Schwinger equation}\label{revisiting}
\label{twosol}
%\alinea 

As was explained in detail in refs.~\cite{Boucaud:2008ky,Boucaud:2010gr,RodriguezQuintero:2010xx}, the low-momentum behavior 
for the Landau gauge ghost dressing function can be inferred from the analysis
of the Dyson-Schwinger equation for the ghost
propagator (GPDSE). That analysis is performed on a very general ground: one applies the MOM renormalization 
prescription,
$F_R(\mu^2) =  \mu^2 \Delta_R(\mu^2) = 1$, 
where $\mu^2$ is the subtraction point, chooses for the ghost-gluon vertex, 
\beq
\widetilde{\Gamma}_\nu^{abc}(-q,k;q-k) &=& 
i g_0 f^{abc} \left( q_\nu H_1(q,k) \right. \nonumber \\
&+& 
\left. (q-k)_\nu H_2(q,k) \right) 
\label{DefH12}
\eeq
to apply this MOM prescription in Taylor kinematics 
({\it i.e.} with a vanishing incoming ghost momentum) 
and assumes the non-renormalizable bare ghost-gluon form factor, $H_1(q,k)=H_1$, 
to be constant in the low-momentum regime for the incoming ghost.
Then, the low momentum-behaviour of the ghost dressing function is 
supposed to be well described by 
%--------
\beq\label{dress}
F_R(q^2) &=& A(\mu^2) \left( \frac{q^2}{M^2} \right)^{\alpha_F} \left( 1 + \cdots 
\rule[0cm]{0cm}{0.6cm} \right) \ ,
\eeq
%--------
and that of the gluon propagator by
%-----------------
\beq\label{gluonprop}
\Delta_R(q^2) &=& \frac{B(\mu^2)}{q^2 + M^2} \ 
\simeq \frac{B(\mu^2)}{M^2} \left( 1 - \frac{q^2}{M^2} + \cdots \right) \ ,
\eeq
%-----------------
and this, after solving asymptotically the GPDSE, finally left us with:
%-----------------
\beq \label{solsFsneq}
F_R(q^2) \simeq 
\left(
\frac {10 \pi^2}{N_C H_1 g_R(\mu^2) B(\mu^2)} 
\right)^{1/2}
\ \left(\frac {M^2} {q^2} \right)^{1/2} \ ,
\eeq
if $\alpha_F \neq 0$; and 
%--------------
\beq\label{solsFseq}
F_R(q^2) &\simeq&
F_R(0) \left( 1 +   
\frac{N_C H_1}{16 \pi} \ \overline{\alpha}_T(0) \ 
\frac{q^2}{M^2} \left[ \ln{\frac{q^2}{M^2}} - \frac {11} 6 \right] \right.
\nonumber \\
&+& \left. {\cal O}\left(\frac{q^4}{M^4} \right) \right)
\eeq
%---------------
if $\alpha_F = 0$, 
where 
%---------------
\beq\label{coefC2}
\overline{\alpha}_T(0) =  M^2 \frac{g^2_R(\mu^2)}{4 \pi} 
F_R^2(0) \Delta_R(0) \ .
\eeq
%---------------
It should be understood that the subtraction momentum for all the renormalization quantities is $\mu^2$. 
The case $\alpha_F \neq 0$ leads to the so-called scaling solution, where the low-momentum behavior of 
the massive gluon propagator forces the ghost dressing function to diverge at low-momentum 
through the scaling condition: $2 \alpha_F + \alpha_G=0$ ($\alpha_G=1$ is the power 
exponent when dealing with a massive gluon propagator). As this scaling condition is verified, 
the perturbative strong coupling defined in this Taylor scheme~\cite{Boucaud:2008gn}, 
$\alpha_T=g_T^2/(4\pi)$, has to reach a constant at zero-momentum,
\beq
\alpha_T(0) &=& \frac{g^2(\mu^2)}{4 \pi} \lim_{q^2\to 0} q^2 \Delta(q^2) F^2(q^2) \nonumber \\
&=& \frac{5 \pi}{2 N_C H_1} \ ,
\eeq
as can be obtained from Eqs.(\ref{gluonprop},\ref{solsFseq}). 
The case $\alpha_F=0$ corresponds to the so-called decoupling solution, 
where the zero-momentum ghost dressing function reaches a non-zero finite value 
and \eq{solsFseq} provides us with the first asymptotic corrections to this 
leading constant. This subleading correction is controlled by the zero-momentum value of 
the coupling defined in \eq{coefC2}, which is an extension of the non-perturbative effective 
charge definition from the gluon propagator~\cite{Aguilar:2008fh} to the Taylor 
ghost-gluon coupling~\cite{Aguilar:2009nf}. 
%As a consequence of the appropriate {\it amputation} of a 
%massive gluon propagator, where the gluon mass scale is the same RI-invariant 
%mass scale appearing in \eq{gluonprop}, 
%this Taylor effective charge is frozen at low-momentum and gives 
%a non-vanishing zero-momentum value.

\section{Numerical results from coupled PT-BFM DSE's}

In ref.~\cite{RodriguezQuintero:2010xx}, the solutions of the coupled DSE system 
in the PT-BFM scheme (with $H_1=1$ for the ghost-gluon vertex), 
numerically integrated for many values of the coupling at 
the renormalization point $\mu^2$ as a boundary condition, 
were studied in the light of the analytical results above presented. 
Here we will shortly discuss the results of this work.

\subsection{The ``regular'' or ``decoupling'' solutions}

\begin{figure}[hbt!]
%\begin{center}
\begin{tabular}{cc}
\includegraphics[width=8cm]{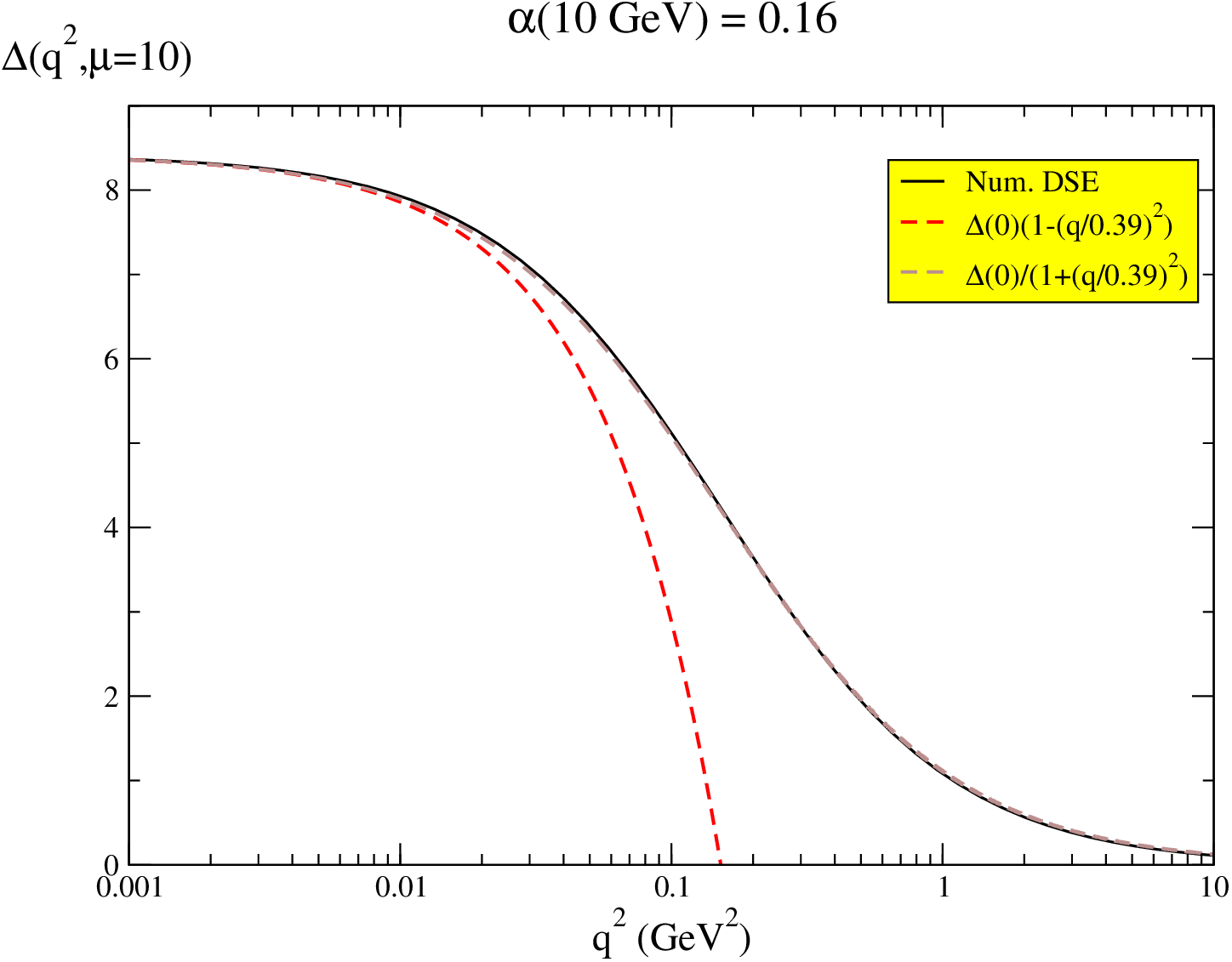} &
\includegraphics[width=8cm]{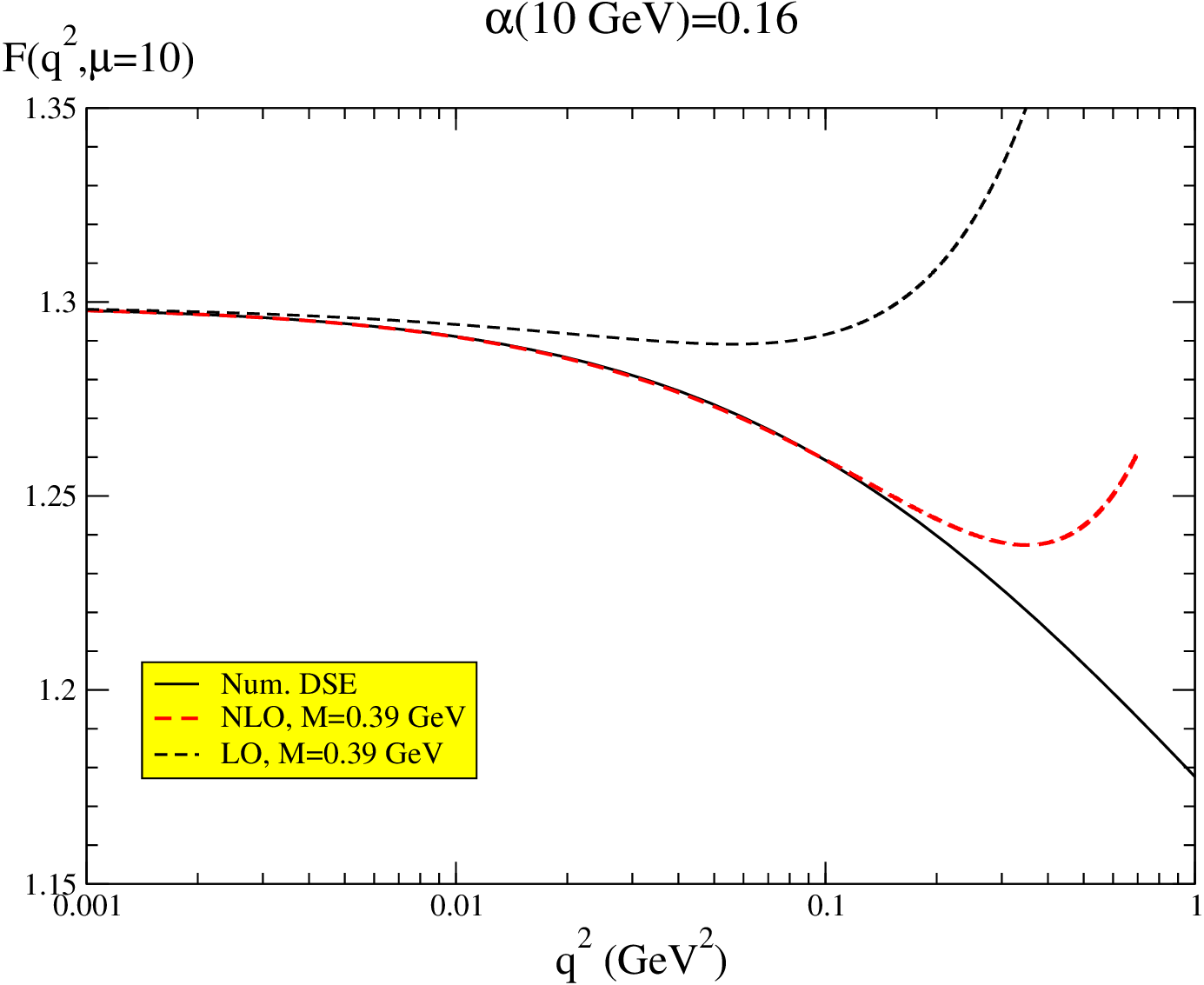} \\
\includegraphics[width=8cm]{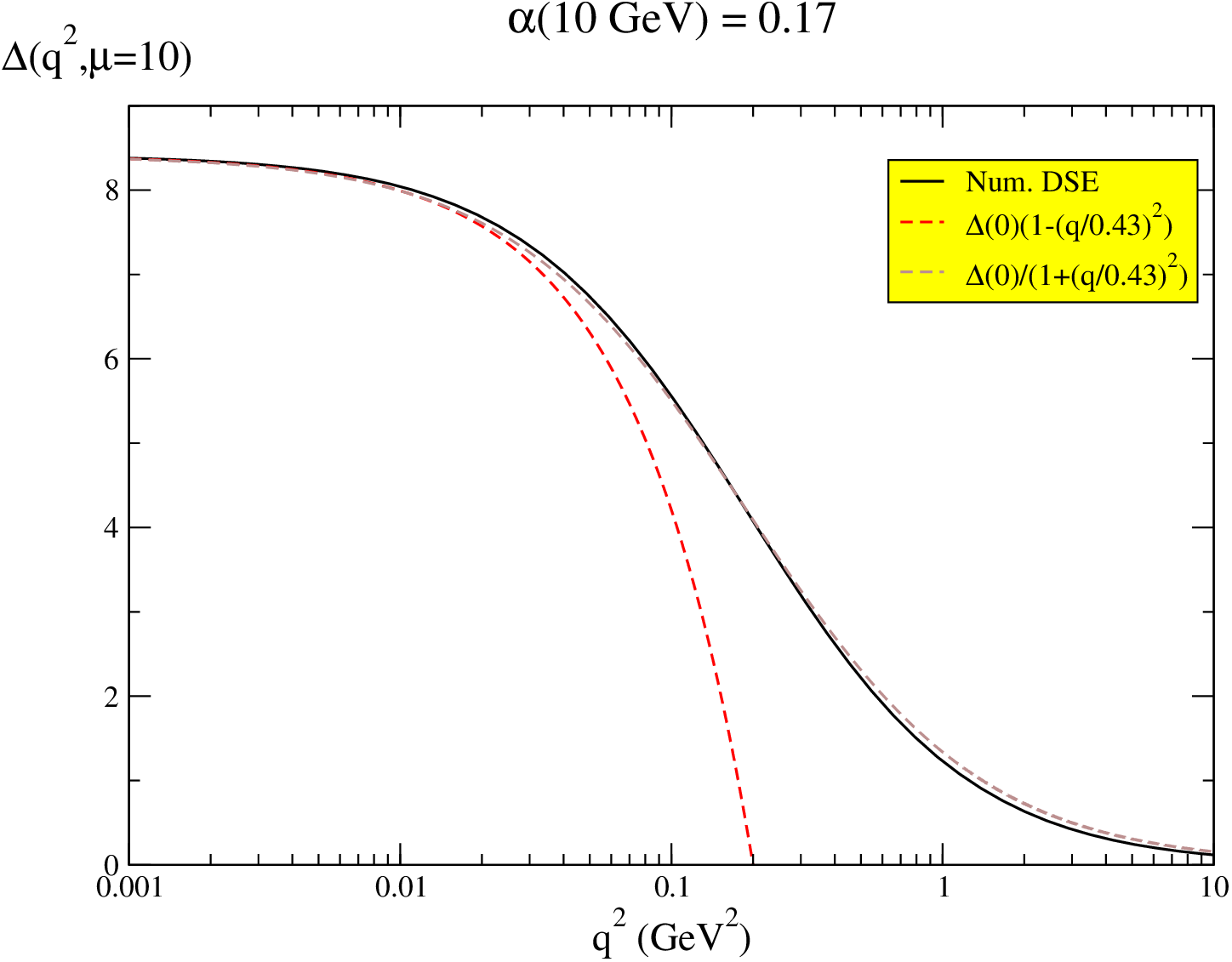} &
\includegraphics[width=8cm]{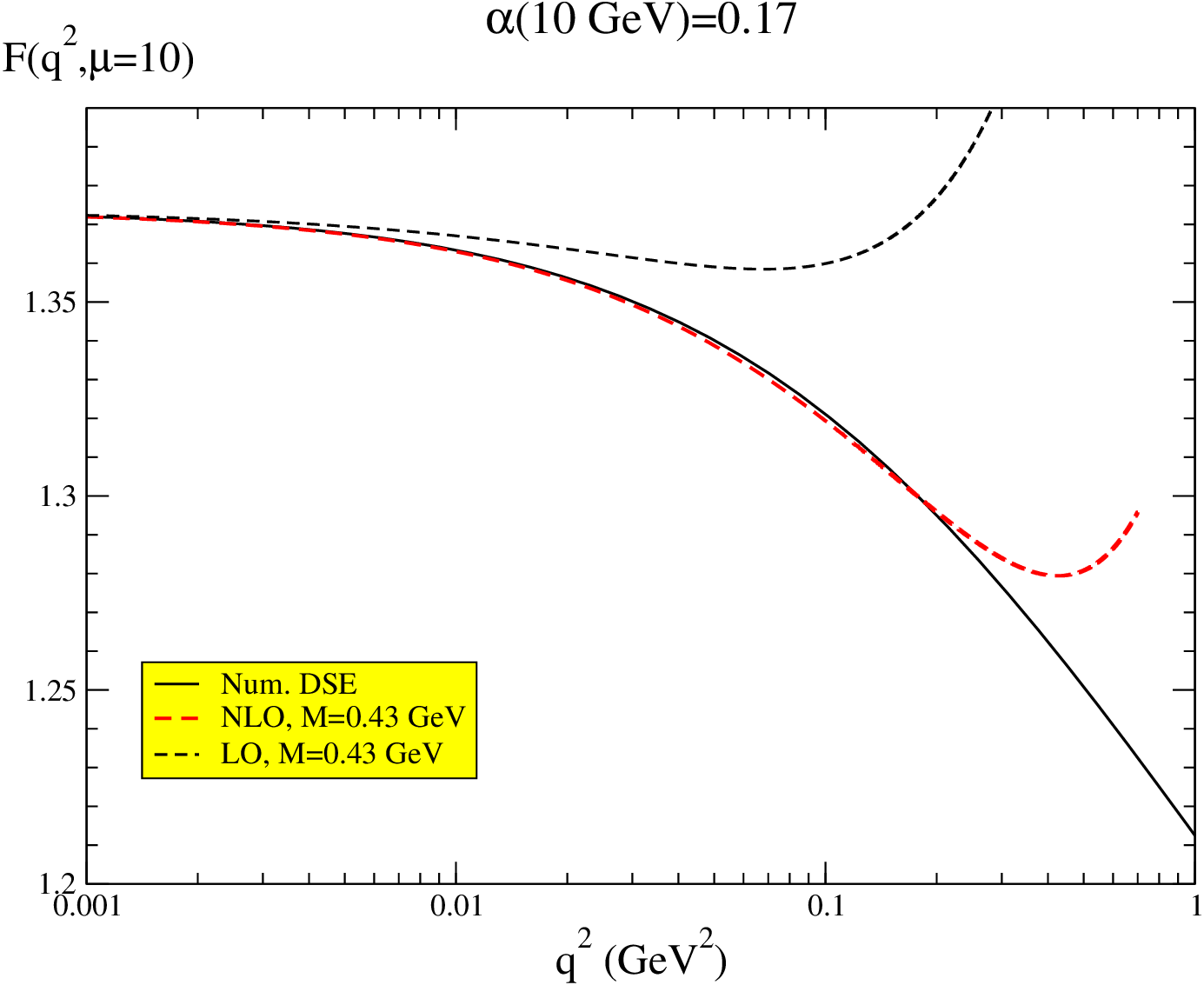} \\
\end{tabular}
%\end{center}
\caption{\small Gluon propagator (left) and ghost dressing function (right) after the numerical integration of  
the coupled DSE system for $\alpha(\mu=10 \mbox{\rm GeV})=0.16,0.17$ taken from \cite{RodriguezQuintero:2010xx} . The curves 
for the best fits to gluon propagator and ghost propagator data explained in the text appear as red dotted lines. 
the same for the black dotted line in the lefthand plot but retaining only the logarithmic leading term for 
the asymptotic ghost dressing function by dropping the $-11/6$ away. }
\label{fig:ghgl}
\end{figure}

The numerical results of the PT-BFM coupled DSEs were shown in 
ref.~\cite{RodriguezQuintero:2010xx} to behave asymptotically as 
\eq{solsFsneq} predicts for the decoupling DSE solutions. 
Indeed, as the gluon propagator solutions in the PT-BFM scheme result to behave as massive ones, 
the eqs.~(\ref{gluonprop},\ref{solsFseq}) must account for the low-momentum behaviour of 
both gluon propagator and ghost dressing function with $H_1=1$ and 
$\overline{\alpha}_T(0)$ given by \eq{coefC2}, with $\alpha_T(\mu^2)=g_R^2(\mu^2)/(4\pi)$ being fixed, 
as a boundary condition for the numerical integration of the coupled DSE for each particular 
solution of the family (see tab. \ref{tab-fits}). Furthermore, the zero-momentum values of the 
ghost dressing function, $F_R(0)$ and of the 
gluon propagator, $\Delta_R(0)$, can be taken from the numerical solutions of the DSE (for any value of the 
$\alpha(\mu=10 \mbox{\rm GeV})$). These altoghether with the gluon masses obtained by the fit of 
\eq{gluonprop} to the numerical DSE gluon propatator solutions  (see 
the left plot in fig.~\ref{fig:ghgl}, for $\alpha(\mu)=0.16$, and the results for $\alpha(\mu)=0.15, 0.16, 0.17$ in 
tab.~\ref{tab-fits}, taken from ref.~\cite{RodriguezQuintero:2010xx}), provide us with all the ingredients 
to evaluate, with no unknown parameter, \eq{solsFseq}. 

%%%%%%%%%%%%%%%%%%%%%%%%%%
%%%%%%%%%%%%%%%%%%%%%%%%%%

Indeed, the expression given by \eq{solsFseq} can be succesfully
applied to describe the solutions all over the range of coupling values, 
$\alpha(\mu)$, at $\mu=10$ GeV (provided that they are not very close of the critical 
coupling that will be defined in the next subsection). 
This can be seen, for instance, for $\alpha=0.16,0.18$, 
in the right plots of fig.~\ref{fig:ghgl} and it was also shown for $\alpha=0.15$ in 
ref.~\cite{RodriguezQuintero:2010xx}.

%%%%%%%%%%%%%%%%%%%%%%%%%%
\begin{table}
%\begin{center}
\begin{tabular}{|c||c|c|} 
\hline
\rule[0cm]{0cm}{0.5cm} $\alpha(\mu)$ & $\overline{\alpha}_T(0)$ & $M$ (GeV) [gluon] \\
\hline
\hline
0.15 & 0.24 & 0.37 
\\
\hline
0.16 & 0.30 & 0.39 \\
\hline
0.17 & 0.41 & 0.43 \\
\hline
\end{tabular}
%\end{center}
\caption{\small Gluon masses and the zero-momentum non-perturbative effective 
charges, taken from ref.~\cite{RodriguezQuintero:2010xx} and obtained as discussed in 
the text.
}
\label{tab-fits}
\end{table}
%%%%%%%%%%%%%%%%%%%%%%%%%%%

\subsection{The ``critical'' limit}

\begin{figure}[!hbt]
%\begin{center}
\begin{tabular}{cc}
\includegraphics[width=8cm]{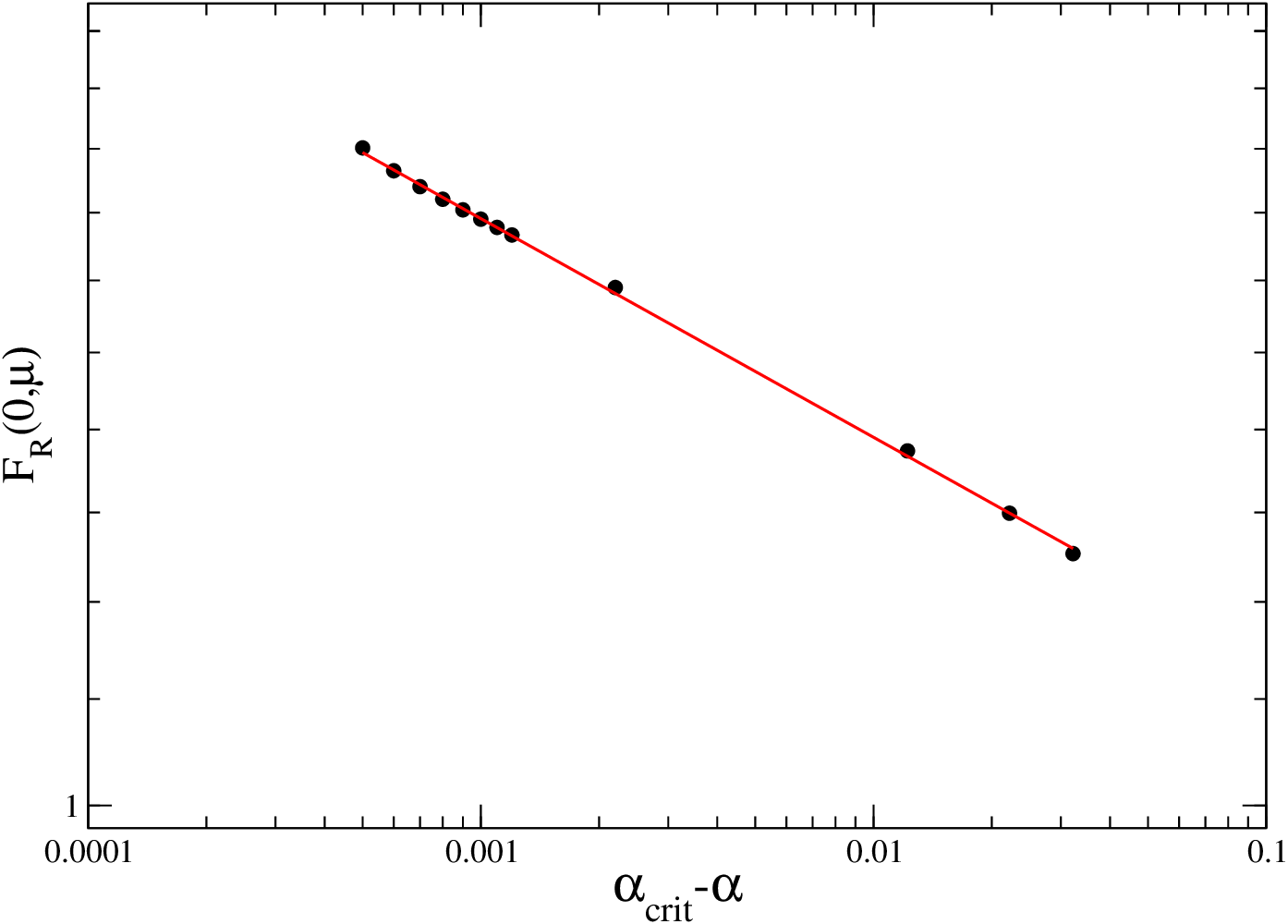} 
&
\includegraphics[width=8cm]{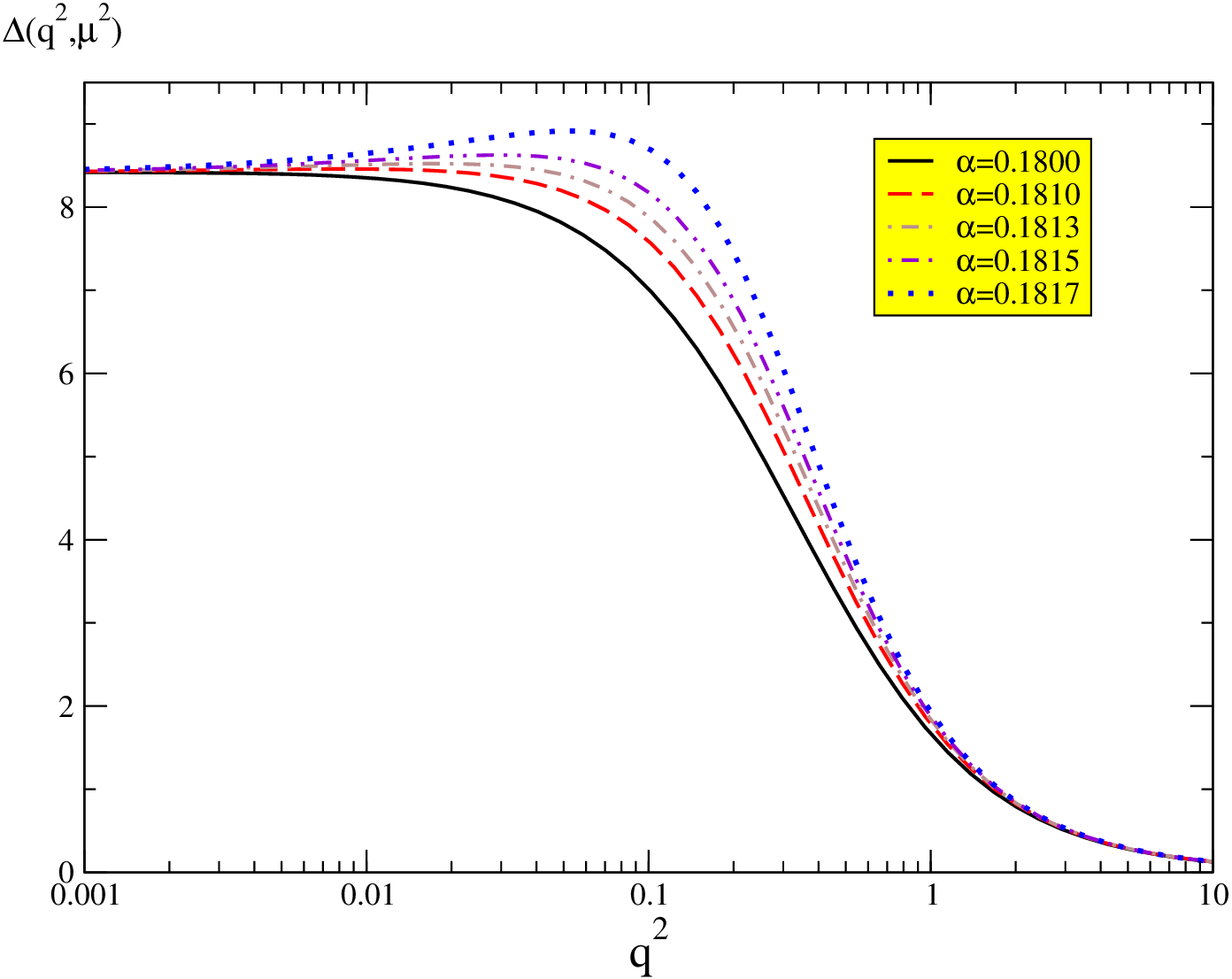} 
\\
(a) & (b) 
\end{tabular}
%\end{center}
\caption{\small (a) Log-log plot of the zero-momentum values of the ghost dressing function, obtained 
by the numerical integration of the coupled DSE system in the PT-BFM scheme, in terms of 
$\alpha_{\rm crit}-\alpha$. $\alpha=\alpha(\mu=10 \mbox{\rm GeV})$, 
the value of the coupling at the renormalization momentum, is an initial condition for 
the integration; while $\alpha_{\rm crit}$ is fixed to be 0.1822, as explained in the text, 
by requiring the best linear correlation. %The negative value for the slope indicates that the 
%zero-momentum ghost propagator diverges as $\alpha \to \alpha_{\rm crit}$. 
(b) Gluon propagator solutions
in terms of $q^2$ for the same coupled DSE system for different values of $\alpha(\mu=10 \mbox{\rm GeV})$, 
all very close to the critical value, ranging from 0.18 to 0.1817~.}
\label{fig:ghost0s}
\end{figure}

There also appeared to be a {\it critical} value of the coupling, 
$\alpha_{\rm crit}=\alpha(\mu^2)\simeq 0.182$ with $\mu=10\mbox{\rm ~Gev}$, 
above which the coupled DSE system does not converge any longer to a solution.
As a matter of fact, we know from refs.\cite{Boucaud:2008ky,RodriguezQuintero:2010xx} that
the scaling solution implies for the coupling
\beq\label{ap:crit}
\alpha_{\rm crit} \ = \ \frac{g_R^2(\mu^2)}{4 \pi} 
\simeq
\frac{5 \pi}{2 N_C A^2(\mu^2) B(\mu^2) } \ ,
\eeq
where $B(\mu^2)$ and $A(\mu^2)$ defined by Eqs.~(\ref{gluonprop},\ref{solsFsneq}).
This is also shown in ref.~\cite{Boucaud:2008ji}, where only the ghost propagator 
DSE with the kernel for the gluon loop integral is obtained from gluon propagator lattice 
data. In the analysis of ref.~\cite{Boucaud:2008ji}, a ghost dressing 
function solution diverging at vanishing momentum appears to exist and 
verifies eqs.~(\ref{solsFsneq},\ref{ap:crit}), while regular or decoupling solutions
 exist for any $\alpha < \alpha_{\rm crit}$. In ref.~\cite{RodriguezQuintero:2010xx}, 
 a more complete analysis is performed: first by studying the solutions for many different values 
 of the coupling, $\alpha=\alpha(\mu^2)$, of a coupled DSE system; and then by showing that 
 the ghost dressing function at vanishing momentum, $F(0,\mu^2)$, 
is described by the following power behaviour,
\beq
F(0) \ \sim \ (\alpha_{\rm crit} - \alpha(\mu^2))^{- \kappa(\mu^2)} \ ,
\eeq
where $\kappa(\mu^2)$ is a critical exponent (depending presummably on the 
renormalization point, $\mu^2$), supposed to be positive and to govern the transition 
from decoupling ($\alpha < \alpha_{\rm crit}$) to the scaling ($\alpha = \alpha_{\rm crit}$) 
solutions; and where we let $\alpha_{\rm crit}$ be a free parameter to be fitted 
by requiring the best linear correlation for $\log[F(0)]$ in terms of 
$\log[\alpha_{\rm crit}-\alpha]$. In doing so, the best correlation 
coefficient is 0.9997 for $\alpha_{\rm crit}=0.1822$, which is pretty close to 
the critical value of the coupling above which the coupled DSE system does not converge 
any more, and $\kappa(\mu^2) = 0.0854(6)$. This can be seen in fig.~\ref{fig:ghost0s}, where 
the log-log plot of $F_R(0)$ in terms of $\alpha_{\rm crit}-\alpha$ is shown and 
the linear behaviour with negative slope corresponding to the best correlation coefficient
strikingly indicates a zero-momentum ghost propagator diverging 
as $\alpha \to \alpha_{\rm crit}$. Nevertheless, no critical or scaling solution 
appears for the coupled DSE system in the PT-BFM, although the decoupling solutions obtained for any 
$\alpha < \alpha_{\rm crit} = 0.1822$ seem to approach the behaviour of a scaling one 
when $\alpha \to \alpha_{\rm crit}$. This is again well understood in ref.~\cite{RodriguezQuintero:2010xx},  
where the gluon propagators obtained 
from the coupled DSE system in PT-BFM were also found to obey the 
same critical behaviour pattern as the ghost propagator, 
when approaching the critical value of the coupling. Indeed, 
when approaching the critical value of the coupling, the gluon propagators obtained 
from the coupled DSE system in PT-BFM must be also thought to obey the 
same critical behaviour pattern as the ghost propagator. 
In the PT-BFM, the value at zero-momentum being 
fixed by construction~\cite{Aguilar:2008xm}, one should expect that, 
instead of decreasing, the 
gluon propagator obtained for couplings near to the critical value increases for low momenta: 
the more one approaches the critical coupling the more it has to increase. 
This is indeed the case, as can be seen in fig.~\ref{fig:ghost0s}(b). 
This also implies that, near the critical value, the low momentum propagator does not 
obey \eq{gluonprop} and that consequently \eq{solsFseq} does not work any longer to 
describe the low momentum ghost propagator.

%%%%%%%%%%%%%%%%%%%%%%%%%
\vspace{0.9cm}
%%%%%%%%%%%%%%%%%%%%%%%%%

\subsection{The critical value and the QCD coupling in the ``real world'' }

Finally, one can pay attention to the critical value of the coupling, $\alpha_{\rm crit}=0.1822$, 
and try to make a comparison with the physical strong coupling values in order get some idea of 
whether the current data can exclude or not this critical behaviour. Although the experimental PDG {\it world 
average} of the strong coupling in the $\overline{\rm MS}$ scheme, $\alpha_{\overline{\rm MS}}(M_Z)=0.1184(7)$~\cite{PDG}, 
can be propagated from the $Z^0$ boson mass down to $\mu=10$ GeV 
to give $\alpha_{\overline{\rm MS}}(10 \ {\rm GeV})=0.179(2)$, that 
incidentally lies on the right ballpark of the above critical value, such a comparison 
is meaningless because our coupling corresponds to one in MOM Taylor-scheme for 
zero number of flavours. One can use instead the available perturbative four-loop formula describing 
the running of the coupling in Taylor-scheme to estimate $\Lambda_{\rm QCD}$ in this particular scheme, 
then perform the conversion to $\overline{\rm MS}$ (see for instance eqs.(22,23) 
of the first reference in~\cite{Boucaud:2008gn}) and thus obtain the value quoted in tab.~\ref{tabLambda}. 
Of course, it would be again meaningless to compare this last value with the one 
for $\Lambda_{\overline{\rm MS}}$ that can be 
obtained from the PDG value for $\alpha_{\overline{\rm MS}}(M_Z)$, also quoted in tab.~\ref{tabLambda}, 
but we can refer the comparison to the lattice Yang-Mills determinations of the same parameter, 
as for instance the two of them included in tab.~\ref{tabLambda}. 
It should be noted that the procedures for the lattice determination of $\Lambda_{\overline{\rm MS}}$ mainly work in 
the UV domain, where IR sources of uncertainties as the Gribov ambiguity or volume effects are indeed 
negligible~\footnote{As a matter of fact, there are unquenched lattice determinations with $N_f=5$ 
staggered fermions for the strong coupling~\cite{Davies:2008sw} which are pretty consistent with the PDG value. 
This can be taken as a good indication in favour of the robustness of the lattice determinations 
of $\Lambda_{\overline{\rm MS}}$.}. 
Thus, the lattice estimates of $\Lambda_{\overline{\rm MS}}$ appear to lie clearly below this critical 
limit for the PT-BFM DSE in pure Yang-Mills. This last results appears to indicate 
that the critical solution is not the one chosen by the zero-flavour world. 
However, as no quark flavour loops effect have been incorporated in our   
DSE analysis, we cannot yet neither compare with the physical strong coupling nor conclude 
whether the critical limit can be allowed in the ``{\it real world}''.

%------------------------------------------------
\begin{table}[htb!]
%\begin{center}
\begin{tabular}{|c|c|c|c|}
\hline 
$\Lambda_{\overline{\rm MS},{\rm crit}}^{N_f=0}$ 
& 
$\Lambda_{\overline{\rm MS}}^{N_f=0}$~\cite{Luscher:1993gh} 
& 
$\Lambda_{\overline{\rm MS}}^{N_f=0}$~\cite{Boucaud:2008gn}
& 
$\Lambda_{\overline{\rm MS}}^{N_f=5}$~\cite{PDG} 
\\
\hline
434 MeV & 238(19) MeV & 244(8) MeV & 213(9) MeV \\
\hline
\end{tabular}
%\end{center}
\label{tabLambda}
\caption{\small The critical value of $\Lambda_{\overline{\rm MS}}$ in pure Yang-Mills inferred 
from $\alpha_{\rm crit}=0.1822$ (first column), lattice estimates for Yang-Mills $\Lambda_{\overline{\rm MS}}$ 
taken from literature (second and third columns) and the one obtained from the PDG value 
of $\alpha_{\overline{\rm MS}}(M_Z)$ by applying a four-loop perturbative formula for the running of  
$\alpha_{\overline{\rm MS}}$ with $N_f=5$.} 
\end{table}
%-------------------------------------------------

\section{Conclusions}\label{conclu}
%\alinea

The ghost propagator DSE, with the only assumption of taking $H_1(q,k)$ from the 
ghost-gluon vertex in \eq{DefH12} to be constant in the infrared domain of $q$, can be 
exploited to look into the low-momentum behaviour of the ghost propagator.  The
two classes of solutions named ``decoupling'' and ``scaling'' can be indentified and 
shown to depend on whether the ghost dressing function achieves a finite non-zero
constant ($\alpha_F=0$) at vanishing momentum or not ($\alpha_F \neq 0$). The 
solutions appear to be dialed by the size of the coupling at the renormalization 
momentum which plays the role of a boundary condition for the DSE integration. 
The low-momentum behaviour of the decoupling solutions results to be regulated by the 
gluon propagator mass and by a regularization-independent 
dimensionless quantity that appears to be the effective charge defined from 
the Taylor-scheme ghost-gluon vertex at zero momentum.

In this note, we have shortly discussed the results of ref.~\cite{RodriguezQuintero:2010xx} 
where the solutions of coupled ghost and gluon propagator DSE in 
the PT-BFM scheme were studied and demonstrated that the asymptotic decoupling formula ($\alpha_F=0$) 
successfully describes the low-momentum ghost propagator. 
The model applied for the massive gluon propagator is also verified to give properly account 
of the gluon solution, at least for momenta below 1 GeV (and for a coupling 
not very close to the critical point).
Although we argued that a massive gluon propagator implies that the ghost dressing 
function takes a non-zero finite value at vanishing momentum, 
we also show that the zero-momentum ghost dressing function 
tends to diverge when the value of the coupling dialing the solutions 
approaches some critical value. Such a divergent behaviour at the critical coupling 
seems to be the expected one for a scaling solution (where, if the gluon 
is massive, $\alpha_F=-1/2$). If we consider the zero-momentum value of the ghost 
dressing function as some sort of ``{\it order parameter}'' indicating whether the 
ghost propagator low-momentum behaviour is suppressed ($\alpha_F=0$ and finite ghost 
dressing function) or it is enhanced ($\alpha_F<0$ and divergent ghost dressing 
function), the strength of the coupling computed at some renormalization point 
seems to control some sort of transition from the {\it suppressed} to 
the {\it enhanced} phases for the ghost propagator DSE solutions in the PT-BFM scheme.
The last only takes place as some critical value of the coupling is reached. 
Neverteless, it can be proven that, as far as the gluon is massive,  
the scaling behaviour for the Yang-Mills propagators appear not to be a 
solution but an unattainable limiting case for the PT-BFM DSE solutions.
 
Finally, the critical value for the coupling is shown to lie clearly much above the 
estimate of the QCD coupling in pure Yang-Mills computed from lattice QCD. This of course 
agrees with the fact that the current large-volume quenched lattice results for ghost 
and gluon Green functions clearly behave as expected for the decoupling solutions.

%%%%%%%%%%%%%%%%%%%%%%%%%%%%%%%%%%%%%%%%%%%%%%%%
%% BACKMATTER
%%%%%%%%%%%%%%%%%%%%%%%%%%%%%%%%%%%%%%%%%%%%%%%%

\begin{theacknowledgments}

The author is particularly indebted to Ph.~Boucaud, J.P~Leroy, A.~Le~Yaouanc, J.~Micheli and 
O.~P\`ene for very fruitful discussions at the initial stages of the work and to 
J.~Papavassiliou and A.C.~Aguilar also for very valuable discussions and comments, and 
specially for providing me with some unpublished results 
which were exploited in this paper. J. R-Q also acknowledges the Spanish MICINN for the 
support by the research project FPA2009-10773 and ``Junta de Andalucia'' by P07FQM02962.

\end{theacknowledgments}



\bibliographystyle{aipproc}   % if natbib is available

%\bibliography{sample}

\begin{thebibliography}{9}

%\cite{Boucaud:2008ji}
\bibitem{Boucaud:2008ji}
  Ph.~Boucaud, J.~P.~Leroy, A.~L.~Yaouanc, J.~Micheli, O.~Pene and J.~Rodriguez--Quintero,
    %``IR finiteness of the ghost dressing function from numerical resolution of
      %the ghost SD equation,''
  JHEP {\bf 0806} (2008) 012
   arXiv:0801.2721 [hep-ph].
     %%CITATION = ARXIV:0801.2721;%%

%\cite{Boucaud:2008ky}
\bibitem{Boucaud:2008ky}
  Ph.~Boucaud, J.~P.~Leroy, A.~Le Yaouanc, J.~Micheli, O.~Pene and J.~Rodriguez-Quintero,
  %``On the IR behaviour of the Landau-gauge ghost propagator,''
  JHEP {\bf 0806} (2008) 099
  [arXiv:0803.2161 [hep-ph]].
  %%CITATION = JHEPA,0806,099;%%
  
%\cite{Aguilar:2006gr}
\bibitem{Aguilar:2006gr}
  A.~C.~Aguilar and J.~Papavassiliou,
  %``Gluon mass generation in the PT-BFM scheme,''
  JHEP {\bf 0612} (2006) 012;
  %[arXiv:hep-ph/0610040].
  %%CITATION = JHEPA,0612,012;%%
%\cite{Aguilar:2007fe}
%\bibitem{Aguilar:2007fe}
%  A.~C.~Aguilar and J.~Papavassiliou,
  %``On dynamical gluon mass generation,''
  Eur.\ Phys.\ J.\  A {\bf 31} (2007) 742;
  %[arXiv:0704.2308 [hep-ph]].
  %%CITATION = EPHJA,A31,742;%%
A.~C.~Aguilar and A.~A.~Natale,
  %``A dynamical gluon mass solution in a coupled system of the  Schwinger-Dyson
  %equations,''
  JHEP {\bf 0408} (2004) 057.
  %[arXiv:hep-ph/0408254].

%\cite{Aguilar:2008xm}
\bibitem{Aguilar:2008xm}
  A.~C.~Aguilar, D.~Binosi and J.~Papavassiliou,
  %``Gluon and ghost propagators in the Landau gauge: Deriving lattice results
  %from Schwinger-Dyson equations,''
  Phys.\ Rev.\  D {\bf 78} (2008) 025010
  [arXiv:0802.1870 [hep-ph]].
  %%CITATION = PHRVA,D78,025010;%%

\bibitem{Alkofer:2000wg}
  R.~Alkofer and L.~von Smekal,
  %``The infrared behavior of QCD Green's functions: Confinement, dynamical
  %symmetry breaking, and hadrons as relativistic bound states,''
  Phys.\ Rept.\  {\bf 353} (2001) 281
  [arXiv:hep-ph/0007355];
  C.~Lerche and L.~von Smekal, 
  %``On the infrared exponent for gluon and ghost propagation in Landau gauge QCD ``
  Phys.\ Rev.\ D {\bf 65} (2002) 125006 [arXiv:hep-ph/0202194];

 D.~Zwanziger, %Non-perturbative Landau gauge and infrared critical exponents in QCD
 Phys.\ Rev.\ D {\bf 65} (2002) 094039 [arXiv:hep-th/0109224];

%\cite{Fischer:2002hna}
%\bibitem{Fischer:2002hna}
  C.~S.~Fischer and R.~Alkofer,
  %``Infrared exponents and running coupling of SU(N) Yang-Mills theories,''
  Phys.\ Lett.\  B {\bf 536} (2002) 177
  [arXiv:hep-ph/0202202];
  %%CITATION = PHLTA,B536,177;%%

  J.~M.~Pawlowski, D.~F.~Litim, S.~Nedelko and L.~von Smekal,
  %'' Infrared behaviour and fixed points in Landau gauge QCD''
  Phys.\ Rev.\ Lett.\ {\bf 93} (2004) 152002 [arXiv:hep-th/0312324].

%\cite{Huber:2007kc}
%\bibitem{Huber:2007kc}
  M.~Q.~Huber, R.~Alkofer, C.~S.~Fischer and K.~Schwenzer,
  %``The infrared behavior of Landau gauge Yang-Mills theory in d=2, 3 and 4
  %dimensions,''
  Phys.\ Lett.\  B {\bf 659} (2008) 434
  [arXiv:0705.3809 [hep-ph]].
  %%CITATION = PHLTA,B659,434;%%

%\cite{Fischer:2008uz}
\bibitem{Fischer:2008uz}
  C.~S.~Fischer, A.~Maas and J.~M.~Pawlowski,
  %``On the infrared behavior of Landau gauge Yang-Mills theory,''
  Annals Phys.\  {\bf 324} (2009) 2408
  [arXiv:0810.1987 [hep-ph]].
  %%CITATION = APNYA,324,2408;%%

%\cite{Kondo:2009ug}
\bibitem{Kondo:2009ug}
  K.~I.~Kondo,
  %``Kugo-Ojima color confinement criterion and Gribov-Zwanziger horizon
  %condition,''
  Phys.\ Lett.\  B {\bf 678} (2009) 322
  [arXiv:0904.4897 [hep-th]];
  %%CITATION = PHLTA,B678,322;%%
%\cite{Kondo:2009gc}
%\bibitem{Kondo:2009gc}
%  K.~I.~Kondo,
  %``Infrared behavior of the ghost propagator in the Landau gauge Yang-Mills
  %theory,''
  Prog.Theor.Phys. {\bf 122} (2010) 1455-1475 
  [arXiv:0907.3249 [hep-th]];
    %%CITATION = ARXIV:0907.3249;%%
%\cite{Kondo:2009wk}
%\bibitem{Kondo:2009wk}
%  K.~I.~Kondo,
  %``Decoupling and scaling solutions in Yang-Mills theory with the Gribov
  %horizon,''
  arXiv:0909.4866 [hep-th].
  %%CITATION = ARXIV:0909.4866;%%

%\cite{Dudal:2009xh}
\bibitem{Dudal:2009xh}
  D.~Dudal, S.~P.~Sorella, N.~Vandersickel and H.~Verschelde,
  %``Gribov no-pole condition, Zwanziger horizon function, Kugo-Ojima
  %confinement criterion, boundary conditions, BRST breaking and all that,''
  Phys.\ Rev.\  D {\bf 79} (2009) 121701
  [arXiv:0904.0641 [hep-th]].
  %%CITATION = PHRVA,D79,121701;%%

%\cite{Aguilar:2009pp}
\bibitem{Aguilar:2009pp}
  A.~C.~Aguilar, D.~Binosi and J.~Papavassiliou,
  %``Indirect determination of the Kugo-Ojima function from lattice data,''
  JHEP {\bf 0911} (2009) 066 
  [arXiv:0907.0153 [hep-ph]].
  %%CITATION = ARXIV:0907.0153;%%

%\cite{Boucaud:2009sd}
\bibitem{Boucaud:2009sd}
  Ph.~Boucaud, J.~P.~Leroy, A.~L.~Yaouanc, J.~Micheli, O.~Pene and J.~Rodriguez-Quintero,
  %``Gribov's horizon and the ghost dressing function,''
  Phys.\ Rev.\  D {\bf 80} (2009) 094501 
    [arXiv:0909.2615 [hep-ph]].
  %%CITATION = ARXIV:0909.2615;%%

%\cite{Cucchieri:2007md}
\bibitem{Cucchieri:2007md}
  A.~Cucchieri and T.~Mendes,
  %``What's up with IR gluon and ghost propagators in Landau gauge? A puzzling
  %answer from huge lattices,''
  PoS {\bf LAT2007} (2007) 297
%  [arXiv:0710.0412 [hep-lat]];
  %%CITATION = POSCI,LAT2007,297;%%
%\cite{Cucchieri:2007rg}
%\bibitem{Cucchieri:2007rg}
%  A.~Cucchieri and T.~Mendes,
  %``Constraints on the IR behavior of the gluon propagator in Yang-Mills
  %theories,''
  Phys.\ Rev.\ Lett.\  {\bf 100} (2008) 241601;
%  [arXiv:0712.3517 [hep-lat]];
%%\cite{Cucchieri:2009zt}
%\bibitem{Cucchieri:2009zt}
%  A.~Cucchieri and T.~Mendes,
  %``Landau-gauge propagators in Yang-Mills theories at beta = 0: massive
  %solution versus conformal scaling,''
%  arXiv:0904.4033 [hep-lat];
  %%CITATION = ARXIV:0904.4033;%%

%\cite{Bogolubsky:2007ud}
%\bibitem{Bogolubsky:2007ud}
%\cite{Bogolubsky:2009dc}
%\bibitem{Bogolubsky:2009dc}
  I.~L.~Bogolubsky, E.~M.~Ilgenfritz, M.~Muller-Preussker and A.~Sternbeck,
  %``Lattice gluodynamics computation of Landau gauge Green's functions in the
  %deep infrared,''
  Phys.\ Lett.\  B {\bf 676} (2009) 69;
%  [arXiv:0901.0736 [hep-lat]];
  %%CITATION = PHLTA,B676,69;%%
%\cite{Bogolubsky:2007ud}
%\bibitem{Bogolubsky:2007ud}
  I.~L.~Bogolubsky, E.~M.~Ilgenfritz, M.~Muller-Preussker and A.~Sternbeck,
  %``The Landau gauge gluon and ghost propagators in 4D SU(3) gluodynamics in
  %large lattice volumes,''
  PoS {\bf LAT2007} (2007) 290;
%  [arXiv:0710.1968 [hep-lat]].
  %%CITATION = POSCI,LAT2007,290;%%

%\bibitem{IlgenGrib}
 A.~Sternbeck, E.-M.~Ilgenfritz, M.~M\:uller-Preussker and A.~Schiller,
 %The gluon and ghost propagator and the influence of Gribov copies  \u2022 ARTICLE
Nucl.\ Phys.\ Proc.\ Suppl.\  {\bf 140} (2005) 653;
%  [arXiv:hep-lat/0409125].
%\bibitem{ilgenfritz2}
% A. ~Sternbeck, E.-~M.~ Ilgenfritz, M. ~Mueller-Preussker, A. ~Schiller
  %The influence of Gribov copies on the gluon and ghost propagator
  AIP Conference Proceedings {\bf 756} (2005) 284,
  [arXiv:hep-lat/0412011];

%\cite{Boucaud:2005ce}
%\bibitem{Boucaud:2005ce}
  P.~Boucaud {\it et al.},
  %``The infrared behaviour of the pure Yang-Mills Green functions,''
 [arXiv:hep-ph/0507104 ];

%\cite{Oliveira:2010xc}
%\bibitem{Oliveira:2010xc}
  O.~Oliveira and P.~Bicudo,
  %``Running Gluon Mass from Landau Gauge Lattice QCD Propagator,''
  arXiv:1002.4151 [hep-lat];
  %%CITATION = ARXIV:1002.4151;%%
%\cite{Dudal:2010tf}
%\bibitem{Dudal:2010tf}
%  D.~Dudal, O.~Oliveira and N.~Vandersickel,
  %``Indirect lattice evidence for the Refined Gribov-Zwanziger formalism and
  %the gluon condensate $\braket{A^2}$ in the Landau gauge,''
%  Phys.\ Rev.\  D {\bf 81} (2010) 074505
%  [arXiv:1002.2374 [hep-lat]].
  %%CITATION = PHRVA,D81,074505;%%

%\cite{Bornyakov:2009ug}
%\bibitem{Bornyakov:2009ug}
  V.~G.~Bornyakov, V.~K.~Mitrjushkin and M.~Muller-Preussker,
  %``SU(2) lattice gluon propagator: continuum limit, finite-volume effects and
  %infrared mass scale m_IR,''
  Phys.\ Rev.\  D {\bf 81} (2010) 054503
%  [arXiv:0912.4475 [hep-lat]].
  %%CITATION = PHRVA,D81,054503;%%

%\cite{Cornwall}
\bibitem{Cornwall}
  J.~M.~Cornwall, Phys. Rev. D {\bf 26}, 1453 (1982).

%\cite{Binosi:2009qm}
%\bibitem{Binosi:2009qm}
  D.~Binosi and J.~Papavassiliou,
  %``Pinch Technique: Theory and Applications,''
  Phys.\ Rept.\  {\bf 479} (2009) 1 
%  [arXiv:0909.2536 [hep-ph]].
  %%CITATION = PRPLC,479,1;%%



%\cite{Dudal:2007cw}
\bibitem{Dudal:2007cw}
%  D.~Dudal, S.~P.~Sorella, N.~Vandersickel and H.~Verschelde,
  %``New features of the gluon and ghost propagator in the infrared region from
  %the Gribov-Zwanziger approach,''
%  arXiv:0711.4496 [hep-th];
  %%CITATION = ARXIV:0711.4496;%%
%\cite{Dudal:2008sp}
%\bibitem{Dudal:2008sp}
  D.~Dudal, J.~A.~Gracey, S.~P.~Sorella, N.~Vandersickel and H.~Verschelde,
  %``A refinement of the Gribov-Zwanziger approach in the Landau gauge: infrared
  %propagators in harmony with the lattice results,''
  Phys.\ Rev.\  D {\bf 78} (2008) 065047
  [arXiv:0806.4348 [hep-th]].
  %%CITATION = PHRVA,D78,065047;%%

%\cite{Frasca:2007uz}
\bibitem{Frasca:2007uz}
  M.~Frasca,
  %``Infrared Gluon and Ghost Propagators,''
  Phys.\ Lett.\  B {\bf 670} (2008) 73
%  [arXiv:0709.2042 [hep-th]].
  %%CITATION = PHLTA,B670,73;%%

%\cite{Tissier:2010ts}
\bibitem{Tissier:2010ts}
  M.~Tissier and N.~Wschebor,
  %``Infrared propagators of Yang-Mills theory from perturbation theory,''
  arXiv:1004.1607 [hep-ph].
  %%CITATION = ARXIV:1004.1607;%%



%\cite{Boucaud:2010gr}
\bibitem{Boucaud:2010gr}
  Ph.~Boucaud {\it al.}, %M.~E.~Gomez, J.~P.~Leroy, A.~L.~Yaouanc, J.~Micheli, O.~Pene and J.~Rodriguez-Quintero,
  %``The low-momentum ghost dressing function and the gluon mass,''
Phys.\ Rev.\  D {\bf 82} (2010) 054007
  [arXiv:1004.4135 [hep-ph]]
  %%CITATION = ARXIV:1004.4135;%%

%\cite{RodriguezQuintero:2010xx}
\bibitem{RodriguezQuintero:2010xx}
J.~Rodriguez-Quintero, 
% ``On the massive gluon propagator, the PT-BFM scheme and the low-momentum behaviour of 
% decoupling and scaling solutions''
[arXiv:1005.4598 [hep-ph]].

%%%%% PT-BFM
%\cite{Binosi:2002ft}
\bibitem{Binosi:2002ft}
  D.~Binosi and J.~Papavassiliou,
  %``The pinch technique to all orders,''
  Phys.\ Rev.\  D {\bf 66} (2002) 111901
  [arXiv:hep-ph/0208189];
  %%CITATION = PHRVA,D66,111901;%%
%\cite{Binosi:2007pi}
%\bibitem{Binosi:2007pi}
%  D.~Binosi and J.~Papavassiliou,
  %``Gauge-invariant truncation scheme for the Schwinger-Dyson equations of
  %QCD,''
  Phys.\ Rev.\  D {\bf 77} (2008) 061702
  [arXiv:0712.2707 [hep-ph]];
  %%CITATION = PHRVA,D77,061702;%%
%\cite{Binosi:2007pi}
%\bibitem{Binosi:2007pi}
  D.~Binosi and J.~Papavassiliou,
  %``Gauge-invariant truncation scheme for the Schwinger-Dyson equations of
  %QCD,''
  Phys.\ Rev.\  D {\bf 77} (2008) 061702
  [arXiv:0712.2707 [hep-ph]].
  %%CITATION = PHRVA,D77,061702;%%



%\cite{Boucaud:2008gn}
\bibitem{Boucaud:2008gn}
  Ph.~Boucaud {\it et al.}, %F.~De Soto, J.~P.~Leroy, A.~Le Yaouanc, J.~Micheli, O.~Pene and J.~Rodriguez-Quintero,
  %``Ghost-gluon running coupling, power corrections and the determination of
  %$\Lambda_{\bar {\rm MS}}$,''
  Phys.\ Rev.\  D {\bf 79} (2009) 014508
  [arXiv:0811.2059 [hep-ph]];
  %%CITATION = PHRVA,D79,014508;%%
%\cite{Sternbeck:2007br}
%\bibitem{Sternbeck:2007br}
  A.~Sternbeck, K.~Maltman, L.~von Smekal, A.~G.~Williams, E.~M.~Ilgenfritz and M.~Muller-Preussker,
  %``Running alpha(s) from Landau-gauge gluon and ghost correlations,''
  PoS {\bf LAT2007} (2007) 256
  [arXiv:0710.2965 [hep-lat]].
  %%CITATION = POSCI,LAT2007,256;%%

%\cite{Aguilar:2008fh}
\bibitem{Aguilar:2008fh}
  A.~C.~Aguilar, D.~Binosi and J.~Papavassiliou,
  %``Infrared finite effective charge of QCD,''
  PoS {\bf LC2008} (2008) 050
  [arXiv:0810.2333 [hep-ph]].
  %%CITATION = POSCI,LC2008,050;%%

%\cite{Aguilar:2009nf}
\bibitem{Aguilar:2009nf}
  A.~C.~Aguilar, D.~Binosi, J.~Papavassiliou and J.~Rodriguez-Quintero,
  %``Non-perturbative comparison of QCD effective charges,''
  Phys.\ Rev.\  D {\bf 80} (2009) 085018
    [arXiv:0906.2633 [hep-ph]].
  %%CITATION = ARXIV:0906.2633;%%

%\cite{PDG}
\bibitem{PDG} 
  K. Nakamura  et al. (Particle Data Group), J. \ Phys. \ G {\bf 37} (2010) 075021

%\cite{Davies:2008sw}
\bibitem{Davies:2008sw}
  C.~T.~H.~Davies {\it et al.} [ HPQCD Collaboration ],
  %``Update: Accurate Determinations of alpha(s) from Realistic Lattice QCD,''
  Phys.\ Rev.\  {\bf D78 } (2008)  114507.
  [arXiv:0807.1687 [hep-lat]].

%\cite{Luscher:1993gh}
\bibitem{Luscher:1993gh}
  M.~Luscher, R.~Sommer, P.~Weisz and U.~Wolff,
  %``A Precise determination of the running coupling in the SU(3) Yang-Mills
  %theory,''
  Nucl.\ Phys.\  B {\bf 413} (1994) 481
  [arXiv:hep-lat/9309005].
  %%CITATION = NUPHA,B413,481;%%


\end{thebibliography}

\IfFileExists{\jobname.bbl}{}
 {\typeout{}
  \typeout{******************************************}
  \typeout{** Please run "bibtex \jobname" to optain}
  \typeout{** the bibliography and then re-run LaTeX}
  \typeout{** twice to fix the references!}
  \typeout{******************************************}
  \typeout{}
 }

%%%%%%%%%%%%%%%%%%%%%%%%%%%%%%%%%%%%%%%%%%%
%% The following lines show an example how to produce a bibliography
%% without the help of the BibTeX program. This could be used instead
%% of the above.
%%%%%%%%%%%%%%%%%%%%%%%%%%%%%%%%%%%%%%%%%%%

\end{document}